\documentclass{iaus}
\usepackage{graphicx}
\usepackage{natbib}
\title[$\sim$100 new X-ray binaries in M31] 
{XMM-Newton reveals $\sim$100 new LMXBs in M31 from variability studies}

\author[R. Barnard et al.]   
{R. Barnard$^1$%
 ,
 L. Shaw Greening$^1$, C. Tonkin$^1$, U. Kolb$^1$ \break \and J.P. Osborne$^2$}

\affiliation{$^1$The Open University, Walton Hall, Milton Keynes, MK7 6AA, UK \break email: R.Barnard@open.ac.uk, L.Shaw-Greening@open.ac.uk., U.C.Kolb@open.a.c.uk\\[\affilskip]
$^2$University of Leicester, University Road, Leicester LE1 7RH, UK\break email: julo@star.le.ac.uk\\
}

\pubyear{2005}
\volume{230}  
\pagerange{119--126}
\date{August 9 2005}
\jname{Proceedings Populations of High Energy Sources in Galaxies}
\editors{Evert J.A. Meurs \& Giuseppina Fabbiano}
\begin{document}

\maketitle

\begin{abstract}
We have conducted a survey of X-ray sources in XMM-Newton observations of M31, examining their power density spectra (PDS) and spectral energy distributions (SEDs). Our automated source detection yielded 535 good X-ray sources; to date, we have studied 225 of them. In particular, we examined the PDS because low mass X-ray binaries (LMXBs) exhibit two distinctive types of PDS. At low accretion rates, the PDS is characterised by a broken power law, with the spectral index changing from $\sim$0 to $\sim$1 at some frequency in the range $\sim$0.01--1 Hz; we refer to such PDS as Type A. At higher accretion rates, the PDS is described by a simple power law; we call these PDS Type B. Of the 225 sources studied to date, 75 exhibit Type A variability, and are almost certainly LMXBs, while 6 show Type B but not Type A, and are likely LMXBs. Of these 81 candidate LMXBs, 71 are newly identified in this survey; furthermore, they are mostly found near the centre of M31. Furthermore, most of the X-ray population in the disc are associated with the spiral arms, making them likely high mass X-ray binaries (HMXBs). In general these HMXBs do not exhibit Type A variability, while  many central X-ray sources  (LMXBs) in the same luminosity range do. Hence the PDS may distinguish between LMXBs and HMXBs in this luminosity range. 

\keywords{X-rays: binaries,  galaxies: individual (M31), accretion discs, black hole physics}
\end{abstract}

\firstsection 
\vspace{-.2in}
\section{Introduction}
We present results from a survey of  X-ray point sources in XMM-Newton observations of the Andromeda Galaxy (M31). M31 is the nearest spiral galaxy to our own, at 760 kpc \citep{vdb00}, and its X-ray population is dominated by X-ray binaries. 


Our work differs from previous surveys of M31 in two important ways. Firstly, we made power density spectra (PDS) from  combined EPIC, background subtracted lightcurves of each source. This was made possible for the first time by the unprecedented sensitivity of XMM-Newton, allowing us to detect variability in M31 X-ray sources on $\sim$100 s timescales. Secondly, we  obtained the 0.3--10 keV luminosity of each source from best fit models to their spectral energy distribution (SED); previous works have used an assumed model to convert from count rate to luminosity in all but the brightest few sources.

We have examined four  XMM-Newton observations of the central region of M31, along with one observation each of the North 1, North 2, North 3, South 1 and South 2 fields \citep[see e.g.][]{pie05}. We performed automated source detection on these regions, and found 535 sources that were observed by all three EPIC detectors \citep[see ][for details]{sgtb05}. However, we have not yet studied these sources in a uniform manner.

\vspace{-.2in}
\section{Combining PDS and SED information}

X-ray binaries exhibit many phenomena on time-scales spanning milliseconds to years. Low mass X-ray binaries (LMXBs) are particularly interesting because they exhibit very distinctive power density spectra that depend more on the accretion rate than on the nature of the primary. Both neutron star and black hole LMXBs exhibit PDS that are characterised by a broken power law at low accretion rates; the spectral index changes from $\sim$0 to $\sim$1 at a break frequency in the range $\sim$0.01--1 Hz \citep{vdk94}.  We call this Type A variability \citep{bko04}. At higher accretion rates, the PDS is described by a simple power law over the some frequency range; we refer to this as Type B variability \citep{bko04}. 
Van der Klis (1994) suggested that the transition from Type A to Type B variability occurs at some constant accretion rate, and suggested a transition at $\sim$1\% Eddington. We have empirical evidence that suggests that the transition from Type A to Type B variability occurs at $\sim$10\% Eddington in the 0.3--10 keV band  \citep{bokh04,bko05}. Since the Eddington limit is proportional to the mass of the primary, we expect black hole LMXBs to exhibit Type A variability at higher luminosities than neutron star LMXBs. We assume a maximum neutron star mass of 3.1 M$_{\odot}$ and classify as a possible black hole LMXB any X-ray source that exhibits Type A variability at a 0.3--10 keV luminosity $>$4$\times$10$^{37}$ erg s$^{-1}$.

We also obtained EPIC-pn SEDs for each source in the 0.3--10 keV band, and an equivalent background SED. Response matrices and ancillary response files were generated for each source. The resulting SEDs were grouped for a minimum of 50 count bin$^{-1}$ from sources with $>$500 counts, and for a minimum of 20 count bin$^{-1}$ for SEDs with $<$500 counts. We fitted several spectral models to each SED and obtained the 0.3--10 keV luminosity from the best fit. If the source+background spectrum contained $<$150 counts, and/or the source spectrum contained $<$20 counts, then the luminosity was estimated from the 0.3--10 keV pn intensity; we assumed an absorbed power law model with line of sight absorption equivalent to 10$^{21}$ H atom cm$^{-2}$ and a photon index of 1.7.  

By combining PDS and SED information we were able to classify the X-ray population of M31 to an unprecedented degree; Table~\ref{class} summarises our classification scheme for LMXBs \citep[see e.g.][]{mr05,bko05}. We do not yet have such a  scheme for HMXBs. For luminosity functions of the different regions of M31, see \citet{sgtb05}.

\begin{table*}[!t]
\centering
\caption{\ Classifications of the X-ray population of M31 by power density spectrum (PDS) and spectral energy distribution (SED). The SED is classified by the photon index of the power law component, $\Gamma$, or by the temperature of the thermal component, k$T_{BB}$. }\label{class}
\begin{tabular}{lllll}
\noalign{\smallskip}
\hline
\noalign{\smallskip}

PDS & SED & Classification\\
\noalign{\smallskip}
\hline
\noalign{\smallskip}
Type A & $\Gamma$ $\sim$1.5--2.1 & NS or BH LMXB in low/hard state \\
Type A & $\Gamma$ $\sim$2.4--3.0 & BH LMXB in very high/\\
& & intermediate (steep power law) state\\
Type B & BB dominated & BH LMXB in high/soft state\\
& k$T_{\rm BB}$ $\sim$0.7--2 keV  \\
Type B & Not BB dominated& NS LMXB in high state\\ 
\noalign{\smallskip}
\hline
\noalign{\smallskip}

\end{tabular}
\end{table*}

\begin{figure*}[!t]
\centering
\includegraphics[angle=0,scale=.5]{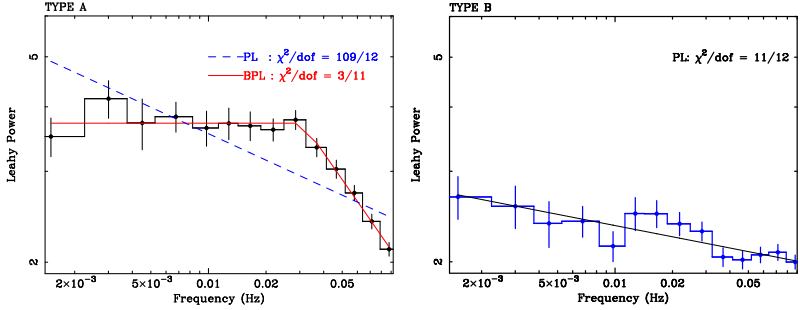}
\caption{ Power density spectra from the 2002 January XMM-Newton observation of two sources in the central region of M31. The PDS are averaged over 96 intervals of 128 bins, lasting 666 s, and grouped geometrically. 
 }\label{pds}
\end{figure*}
\vspace{-.2in}
\section{Results}
The initial survey covered the four observations of the 63 brightest sources in the central region of M31 \citep{bko05}; we have yet to study the $\sim$130 remaining sources. Similarly, we have only surveyed 36 out of 87 sources in South 1 region; this survey also predates the automated source detection.  However, we have studied all 83 sources in North 1 and all 43 sources in North 2. We have not yet studied any of the 88 sources in North 3 nor the 41 sources in South 2. All the remaining sources will be surveyed over the next few months.

We have examined the four observations of the 63 X-ray sources in the core in most detail \citep{bko05}; we were able to classify 107 out of 252 PDS as either Type A, Type B or flat; flat PDS had no significant power above 2, which is the expected power for Poisson noise when using Leahy normalisation.  Figure~\ref{pds} shows examples of Type A and Type B PDS exhibited by M31 X-ray sources during the $\sim$60 ks XMM-Newton observation on 2002, January 6. Furthermore, we found that the Type B and flat PDS were both consistent with  high accretion rate LMXBs in M31 \citep{bko05}.


In the core, 40 out of 63 sources exhibited Type A variability in at least one observation; these are almost certainly X-ray binaries. A further 6 sources exhibited Type B but not Type A variability, making them likely LMXBs. In the North 1 and North 2 fields, 25 out of 83 and 1 out of 43 sources respectively exhibit Type A variability; in the South 1 field, 9 out of 36 sources exhibited Type A variability. Of the 40 sources that exhibited Type A variability in the core, 13 did so at luminosities $>$4$\times$10$^{37}$ erg s$^{-1}$, and were classed as black hole candidates \citep[see ][]{bko05}.

\begin{figure*}[!t]
\resizebox{\hsize}{!}{\includegraphics[angle=0]{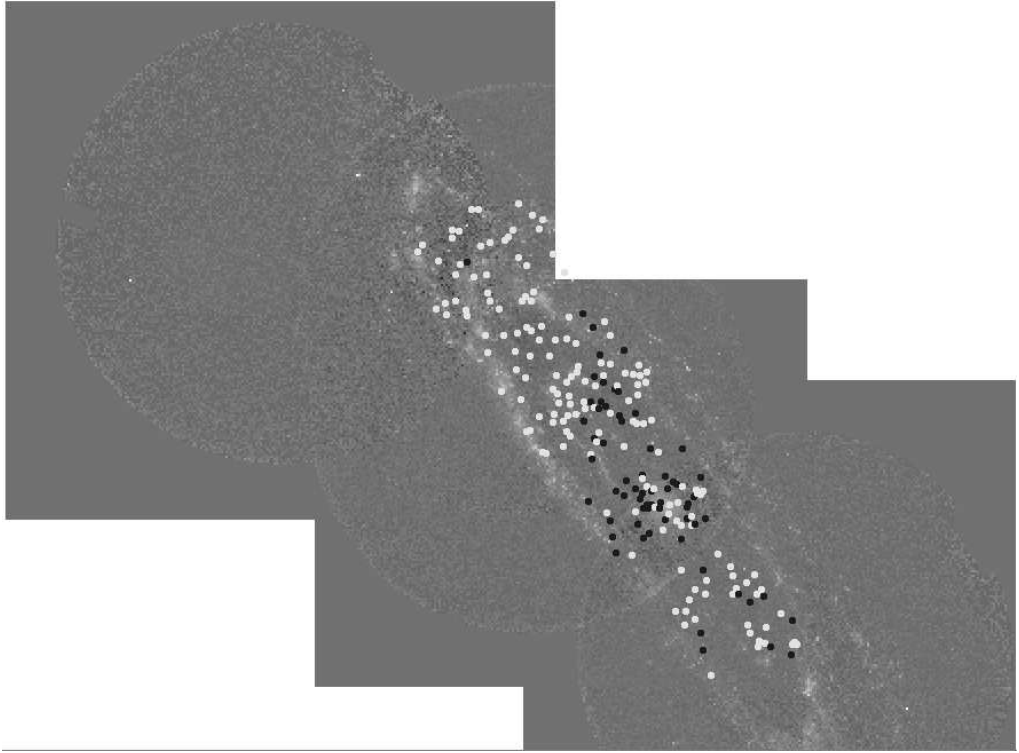}}
\caption{X-ray sources that we have already surveyed, superposed on a mosaic of GALEX FUV observations of M31. Black sources exhibit Type A variability, and are hence likely LMXBs. White sources do not exhibit Type A variability. Out of the 225 sources surveyed to date, 75 exhibit Type A variability. }\label{uvx}
\end{figure*}

\citet{pie05} surveyed these same XMM-Newton observations, finding 856 point sources. They catalogued 7 LMXBs and 9 likely LMXBs, as well as 27 certain and 10 likely associations of X-ray sources with globular clusters. These 53 sources are all likely LMXBs. We have surveyed 28 of these sources to date, and find that 10 of them exhibit Type A variability. Hence, 71 of the 81 X-ray binary candidates were newly identified by this work.         . 
In Fig.~\ref{uvx} we present a mosaic of GALEX images of M31 in the far ultraviolet \citep[see e.g][and http://www.galex.caltec.edu]{thil05}, with the X-ray sources that we have already studied superposed. Black dots represent X-ray sources that exhibited Type A variability in at least one observation, while white dots  represent sources that have not. We note that the central region is dominated by black dots; 63\% of the surveyed sources exhibited Type A variability in the core. Meanwhile, 25--30\% of sources in the North 1 and South 1 regions, and only 2\% of sources in North 2, exhibited Type A variability (see Fig.~\ref{uvx}).  

The majority of sources in the North 1, North 2 and South 1 regions have 0.3--10 keV luminosities $<$10$^{37}$ erg s$^{-1}$. Almost all the core X-ray sources that we have surveyed in this regime with classified PDS show Type A variability; hence we would expect all LMXBs in the M31 disc to exhibit Type A variability also. However, we only detect Type A variability in $\sim$20\% of the disc sources. Furthermore, we see that the disc population of X-ray sources largely follows the spiral arms, suggesting that they are HMXBs. We therefore suggest that it may be possible to distinguish between LMXBs and HMXBs, based on their PDS at 0.3--10 keV luminosities $<$10$^{37}$ erg s$^{-1}$; the  HMXBs are not expected to exhibit Type A variability as they are not disc fed. We expect most of the $\sim$130 faint sources in the Core to exhibit Type A variability.


\vspace{-.1in}
\begin{acknowledgments}
For this work, we used GALEX data that is publicly available from the archive at  http://www.galex.stsci.edu/GR1. RB is supported by PPARC.
\end{acknowledgments}

\end{document}